\documentclass[12pt,preprint]{aastex}

\shorttitle{Short Title} \shortauthors{Fan}

\begin{document}


\title{Intrinsic alignments of galaxies and their effects on 
weak lensing detections of mass concentrations}


\author{Z.-H. Fan}
\affil{Department of Astronomy, Peking University,
    Beijing 100871, China}
\email{fan@bac.pku.edu.cn}




\begin{abstract}

In this paper we investigate the influence of the intrinsic alignment of background galaxies
on weak lensing detections of mass concentrations. Specifically,
we analyze the number counts of false peaks resulting from intrinsic ellipticities
in lensing convergence maps. Including the alignment of source galaxies, 
the full noise variance from intrinsic ellipticites in convergence $\kappa$-maps can be written as 
$\sigma^2_0=\sigma^2_{0ran}+\sigma^2_{0corr}$, where $\sigma^2_{0ran}$
is the noise contributed from randomly oriented source galaxies and
$\sigma^2_{0corr}$ denotes the additional noise from intrinsic alignments.
However, it is observationally difficult to measure $\sigma^2_{0corr}$ 
and usually only $\sigma^2_{0ran}$ can be estimated in weak lensing observations.
Thus the observational signal-to-noise ratio is often defined with respect to $\sigma_{0ran}$,
which is denoted as $\nu_{ran}$ in this paper.  
The true signal-to-noise ratio $\nu$ in terms of $\sigma_0$ is then
$\nu=\nu_{ran}/(1+\sigma^2_{0corr}/\sigma^2_{0ran})^{1/2}$.
Given a detection threshold on $\nu_{ran}$, a larger value of $\sigma^2_{0corr}/\sigma^2_{0ran}$
leads to a lower threshold on $\nu$ and therefore a larger expected number of 
false peaks.  
With $\sigma^2_{0corr}/\sigma^2_{0ran}\sim 10\%$,  
the average number of false peaks with $\nu_{ran}\ge 3.5$ nearly doubles 
compared to that without considering the alignment, 
and for $\nu_{ran}\ge 5$, the number is tripled. 
As a result, the efficiency of weak lensing cluster detection 
degrades significantly. The increase of the number of false peaks 
also affects the likelihood of the existence of dark clumps. 
On the other hand, if one can observationally distinguish 
false peaks and peaks associated with real mass concentrations, e.g., using lensing tomography 
and follow up observations, the number of false peaks can be used to constrain tightly the 
level of intrinsic alignments of source galaxies.  
The CFHTLS Deep $3.61\hbox{ deg}^2$ lensing observations and follow up studies 
find that $5$ out of the $14$ peaks with $\nu_{ran} > 3.5$ are likely to be false peaks, giving
rise to a constraint $\sigma^2_{0corr}\le 1.6\times 10^{-5}$ ($1\sigma$) at the angular
scale of $1\hbox{ arcmin}$ for galaxies at redshift $z\sim 1$. This 
corresponds to $C_{11}+C_{22}\le 3.2\times 10^{-5}$, 
where $C_{11}$ and $C_{22}$ are, respectively, the angular correlations
of intrinsic ellipticities $e_1$ and $e_2$ of background galaxies.  
This result is fully consistent with the limits on the intrinsic alignment derived directly
from observations of Sloan Digtal Sky Servey.

\end{abstract}




%
\section{Introduction}
Gravitational lensing effects are the only ways to directly measure the
distribution of dark matter in the universe (e.g., Hoekstra et al. 2006). 
Because of their dependence on the formation of structures as well as on the geometry of the universe,
lensing effects are sensitive to the nature of dark energy, and
therefore are highly promising in dark energy studies (e.g., Knox et al. 2006; Munshi et al. 2006). 
Weak gravitational lensing effects are mostly extracted from image distortions of 
background galaxies (e.g., Bartelmann \& Schneider 2001).
The intrinsic ellipticities of galaxies therefore present themselves as important 
errors in lensing observations (e.g., Kaiser \& Squires 1993; Schneider 1996). 
It has been commonly assumed that the intrinsic ellipticities of different background galaxies 
are statistically uncorrelated, and thus the average shear measured over a large enough 
number of galaxies gives an unbiased estimate on the lensing effects (e.g., Kaiser \& Squires 1993). 
However, the formation of galaxies is highly affected by their environment, 
and the shapes of galaxies can well be correlated if they are close enough. 
With the assumption that the shapes of galaxies are well represented by the shapes
of their host dark matter halos, numerical simulations indicate that the shape correlations 
range from $10^{-5}$ to $10^{-3}$ on angular scales of a few arcminutes depending on 
halos masses and redshift distributions of source galaxies (e.g., Heavens et al. 2000;
Croft \& Metzler 2000; Jing 2002; Porciani et al. 2002; Heymans et al. 2006). The existence of such 
correlations can contaminate lensing signals significantly. 
Weak lensing effects are directly related to lensing potentials, and thus
only the gradient modes, i.e., $E$-modes, are expected (e.g., Crittenden et al. 2001).  
The presence of $B$-modes can therefore be used to reveal the
existence of different systematics including the intrinsic alignments,
but the correction to the $E$-mode amplitude cannot be done in a straightforward way 
(e.g., Heymans et al. 2004). Down-weighting or removing physically closed pairs 
of background galaxies in lensing analyses reduces the contamination of intrinsic 
alignments at the expense of increasing shot noises if the alignments extend to 
relatively large scales (e.g., Heymans et al. 2004). Because of the different redshift-dependence
for lensing signals and for intrinsic alignments, the tomographic method based on 
template fitting has been proposed to isolate different components assuming the availability
of photometric redshifts for background galaxies (King \& Schneider 2003). This method
has also been extended to include shear-ellipticity cross correlations in the analyses
(e.g., Hirata \& Seljak 2004; King 2006).
 
The intrinsic alignments of galaxies have been searched observationally. The SuperCOSMOS
data on nearby galaxies (with median redshift $z\sim 0.1$) reveal a level of $10^{-5}$ to 
$10^{-4}$ on the correlation of intrinsic ellipticities of galaxies over the angular 
scales of a few tens of arcminutes (Brown et al. 2002). 
Analyses on close pairs of galaxies from COMBO-17 data (with $z\sim 0.6$) 
find the intrinsic alignments to be consistent with zero
but with uncertainties on the order of $a\  few \times 10^{-4}$ 
on scales of a few arcminutes (Heymans et al. 2004). 
Investigations on SDSS main sample with $z\sim 0.1$ and its subsamples
with $z\sim 0.07$ to $z\sim 0.21$ conclude that no significant intrinsic alignments
are detected (Mandelbaum et al. 2006). The observational results
on the intrinsic alignments are close to the lower limits given by different numerical 
simulations on dark matter halos (e.g., Heymans et al. 2006). 
It has been pointed out that the existence of 
misalignment between baryonic matter and dark matter can significantly
reduce the intrinsic alignment of background galaxies in comparison with 
that of dark matter halos and may explain the low observational results 
found in different surveys (Heymans et al. 2006).

In this paper, we study the effects of the intrinsic alignment on finding
mass concentrations through weak lensing effects. Being the largest virialized objects in the universe,
clusters of galaxies are important cosmological probes because their formation
and evolution depend sensitively on cosmologies (e.g., Borgani 2006; Fan \& Chiueth 2001). 
However, large uncertainties exist in linking cluster observables,
such as galaxy richness, X-ray brightness and Sunyaev-Zeldovich effect,
to their mass, the important quantity in cosmological analyses (e.g., Bode et al. 2006).  
On the other hand, lensing effects are generated through gravitation, and depend 
on the total mass distribution.
Thus it is expected that a cluster sample detected through weak lensing effects
is better suited for cosmological studies in comparison with those selected by other probes
(e.g., White et al. 2002; Hamana et al. 2004; Tang \& Fan 2005; Fang \& Haiman 2006).  
Without involving complicated gas physics, weak lensing cluster detections, however, have their own
shortcomings. Besides observational errors, physical systematics, such as
projection effects and complex mass distributions of clusters
of galaxies, affect the selection function of weak lensing clusters considerably.
Thus weak lensing cluster samples are not truly mass-selected (Tang \& Fan 2005).
The intrinsic ellipticities of background galaxies result false peaks in lensing 
maps and reduce the efficiency of cluster detections significantly (e.g., White et al. 2002). 
The false peaks could also be misinterpreted as dark clumps, which might lead to 
a faulty conclusion regarding the validity of a cosmological model. 
Here we explore how the existence of the intrinsic alignment of background galaxies
affects the number of false peaks in lensing convergence maps. We further propose
that the number of false peaks can be used to constrain sensitively the level of the 
intrinsic alignment if one can separate true and false peaks observationally.

The rest of the paper is organized as follows. In \S2, we discuss the 
correlations of galaxy ellipticities based on the model proposed by Heymans et al. (2004;
Heymans et al. 2006).
In \S3, we study the dependence of the number of false peaks on the intrinsic
alignment of background galaxies. In \S4, we analyze the constraints on the intrinsic
alignment from the results of CFHTLS Deep on the number of false peaks given by 
Gavazzi and Soucail (2007). Discussions are presented in \S5.

\section{Intrinsic alignments of galaxies}

Galaxies do not form in isolated ways. Environmental effects 
play important roles in shaping galaxies. Therefore 
correlations of ellipticities of galaxies are expected if they are 
close enough. 



The ellipticity of a galaxy is defined through the second moments of its surface
brightness profile $S(x,y)$. Specifically, we adopt the following definitions
\begin{equation}\label{w1}
2e_1={I_{xx}-I_{yy}\over I_{xx}+I_{yy}},\ \ \ \   2e_2={2I_{xy}\over I_{xx}+I_{yy}},
\end{equation}
where ($I_{yy}$ and $I_{xy}$ have similar forms)
\begin{equation}\label{w1}
I_{xx}={\int S(x,y)(x_i-\bar x)(x_i-\bar x)dx dy\over \int S(x,y) dx dy}.
\end{equation}
Here $(\bar x, \bar y)$ are the coordinates of the center of the galaxy image.
Concerning two-point correlations $c_{ij}(\vec r)=<e_i(\vec x)e_j(\vec x+\vec r)>$, 
it is convenient to choose $x$-axis and $y$-axis
to be parallel and perpendicular to the line joining the two considered galaxies in the
projected plane. 

Numerical simulations show that $c_{12}=<e_1(\vec x)e_2(\vec x+\vec r)>\approx 0$ 
(e.g., Jing 2002; Heymans et al. 2004; Heymans et al. 2006). For 
$c_{ii}=<e_i(\vec x)e_i(\vec x+\vec r)>$ ($i=1,\hbox{ 2}$), 
we use the fitting formula provided by Heymans et al. (2004), which is 
\begin{equation}\label{w1}
c_{ii}={0.001A_i\over 1+(r/B_i)^2}.
\end{equation}
Our following analyses primarily concern $\eta(r)=<e_1(\vec x)e_1(\vec x+\vec r)>
+<e_2(\vec x)e_2(\vec x+\vec r)>=c_{11}+c_{22}$,  which can also be written as 
\begin{equation}\label{w1}
\eta(r)={0.001A\over 1+(r/B)^2}.
\end{equation}
Incorporating different galaxy models in numerical simulations, the fitting values 
of $A$ and $B$ are obtained for each model by Heymans et al. (2004; Heymans et al. 2006).
Comparing with SDSS observations, Mandelbaum et al. (2006) present their fitting results
with $B=1h^{-1}\hbox{Mpc}$, and $A=0.57\pm 0.72$ (see also Heymans et al. 2006).

To investigate their influence on weak lensing effects, we need to analyze the angular
correlation of intrinsic ellipticities, which is related to the three dimensional 
correlation $c_{ij}(\vec r)$ through the following equation
\begin{equation}\label{w1}
C_{ij}(\theta)={\int r_1^2\phi(r_1)r_2^2\phi(r_2)dr_1dr_2[1+\xi(r_{12})]c_{ij}(r_p,\pi) 
\over \int r_1^2\phi(r_1)r_2^2\phi(r_2)dr_1
dr_2[1+\xi(r_{12})]},
\end{equation}
where $\phi(r)$ and $\xi(r)$ are the selection function and the two-point correlation function
for background galaxies, and $r_p$ and $\pi$ are the 
comoving separations of two galaxies perpendicular and along the line of sight, respectively.
Since the correlations decrease quickly on large scales, in the small-angle limit we have
(e.g., Jing 2002),
\begin{equation}\label{w1}
C_{ij}(\theta)={\int r^4\phi(r)^2 dr \Sigma_{ij}(r\theta)
\over [\int r^2\phi(r) dr]^2+\int r^4 \phi(r)^2 dr\int d\pi \xi(r\theta,\pi)},
\end{equation}
where 
\begin{equation}\label{w1}
\Sigma_{ij}(r_p)=\int d\pi [1+\xi(r_p,\pi)]c_{ij}(r_p,\pi).
\end{equation}

%

For weak lensing effects, both the convergence $\kappa$ and the shear $\gamma$
are determined by the second derivatives of the lensing potential $\phi$, and
\begin{equation}\label{w1}
\kappa={\nabla^2 \phi \over 2}, \quad \gamma_1={(\phi_{,11}-\phi_{,22})\over 2},
\quad \gamma_2=\phi_{,12},
\end{equation}
where $\phi_{,ij}=\partial_i\partial_j \phi$.

Concerning weak lensing detections of mass concentrations, we focus on the 
convergence $\kappa$ field.  In the weak lensing limit, it is related to the
shear $\gamma$ in Fourier space through
\begin{equation}\label{w1}
\tilde\kappa(\vec k)=c_{\alpha}(k)\tilde \gamma_{\alpha}(\vec k),
\end{equation}
where the summation over $\alpha=(1,2)$ is implied, and 
$c_{\alpha}=[\cos (2\varphi),\sin (2\varphi)]$ with 
$\vec k=k(\cos \varphi,\sin \varphi)$ (Kaiser \& Squires 1993).
Observationally, the shear $\gamma$ can be estimated from the
ellipticities of galaxy images. In the weak lensing limit, we have
\begin{equation}\label{w1}
\mathbf{e^{(O)}}\approx {\mathbf{\gamma}} + \mathbf{e^{(S)}},
\end{equation}
where $\mathbf{e}$ is defined in eq.(1), and the superscripts 'O' and 'S' denote
observed image and source, respectively. Then the noisy convergence $\kappa_n$ 
including the contamination from source ellipticities follows
\begin{equation}\label{w1}
\tilde \kappa_n(\vec k)=c_{\alpha}(k)[\tilde e^{(O)}_{\alpha}(\vec k)]
=\tilde \kappa(\vec k)+c_{\alpha}(k)[\tilde e^{(S)}_{\alpha}(\vec k)].
\end{equation}
Considering smoothed quantities, we have (e.g., van Waerbeke 2000)
\begin{equation}\label{w1}
\mathbf{\Sigma}^{(O)}(\vec \theta)=\mathbf\Gamma(\vec \theta)
+{1\over n_g}\sum_{i=1}^{N_g} W(\vec \theta -\vec \theta_i)\mathbf{e}^{(S)}(\vec \theta_i),
\end{equation}
and
\begin{equation}\label{w1}
K_N(\vec \theta)=\int d\vec k \hbox{ } e^{-i\vec k \cdot \vec \theta}
c_{\alpha}(k)\tilde \Sigma^{(O)}_{\alpha}(\vec k),
\end{equation}
where $\mathbf{\Sigma}^{(O)}$, $\mathbf{\Gamma}$, and $K_N(\vec \theta)$ are
the smoothed $\mathbf{e^{(O)}}$, $\gamma$ and $\kappa_n$, respectively,
$W(\vec \theta)$ is the smoothing function, and $n_g$ and $N_g$ are,
respectively, the surface number density and the number of source galaxies in the field.
The noise part of $K_N$ due to the intrinsic ellipticities is then
\begin{equation}\label{w1}
N(\vec \theta)={1\over n_g}\sum_{i=1}^{N_g}\int d\vec k \tilde W(\vec k)\hbox{ }
e^{-i \vec {k}\cdot (\vec \theta-\vec \theta_i)}c_{\alpha}(k)e_{\alpha}^{(S)}(\vec \theta_i),
\end{equation}
where $\tilde W(\vec k)$ is the Fourier transformation of the smoothing function with the form
\begin{equation}\label{w1}
\tilde W(\vec k)={1\over (2\pi)^2}\int d\vec \theta \hbox{ }e^{i \vec k \cdot \vec \theta} W(\vec \theta).
\end{equation}
Following van Waerbeke (2000), the correlation of $N(\vec \theta)$ is calculated by averaging
over both the ellipticities and the positions of source galaxies.  Without intrinsic alignments, 
the correlation of $N(\vec \theta)$ arises only from the smoothing operations, and
by ignoring the non-uniform sampling of source galaxies, we have 
(van Waerbeke 2000) 
\begin{equation}\label{w1}
<N(\vec \theta)N(\vec \theta^{\prime})>={\sigma^2_{\epsilon}\over 2 n_g}(2\pi)^2\int d\vec k 
\hbox{ }e^{i \vec {k}\cdot (\vec \theta^{\prime}-\vec \theta)}|\tilde W(\vec k)|^2,
\end{equation}
where $\sigma_{\epsilon}$ is the intrinsic dispersion of $\mathbf{e}^{(S)}$, and the 
factor $(2\pi)^2$ comes in to be in accord with the definition of $\tilde W(\vec k)$ in eq. (15). 

Including the alignment, the operation by averaging over the ellipticities of source galaxies,
denoted by ${\it {A}}$ following van Waerbeke (2000), is
\begin{equation}\label{w1}
{\it {A}} [e^{(S)}_{\alpha}(\vec \theta_i)e^{(S)}_{\beta}(\vec \theta_j)]={\sigma^2_{\epsilon}\over 2}
\delta_{\alpha \beta}\delta(\vec \theta_i-\vec \theta_j)+\delta_{\alpha \beta}C_{\alpha \beta}
(\vec \theta_i-\vec \theta_j),
\end{equation}
where $C_{\alpha \beta}(\vec \theta_i-\vec \theta_j)$ is given in eq. (6). 
Further by averaging over positions of galaxies, i.e., by applying the
operation $(1/S^2)\int d\vec \theta_i d\vec \theta_j$ with $S$ being the area of the 
field (van Waerbeke 2000), we get 
\begin{eqnarray}\label{w11}
<N(\vec \theta)N(\vec \theta^{\prime})>=&&{\sigma^2_{\epsilon}\over 2 n_g}(2\pi)^2\int d\vec k\hbox{ }e^{i \vec {k}\cdot (\vec \theta^{\prime}-\vec \theta)}|\tilde W(\vec k)|^2 \nonumber \\
&& +(2\pi)^4\int d\vec k\hbox{ }e^{i \vec {k}\cdot (\vec \theta^{\prime}-\vec \theta)}|\tilde W(\vec k)|^2[c_1^2(k)\tilde C_{11}(\vec k)+c_2^2(k)\tilde C_{22}(\vec k)],
\end{eqnarray}
where $\tilde C_{11}(\vec k)$ and $\tilde C_{22}(\vec k)$ are the corresponding Fourier
transformations of $C_{11}(\theta)$ and $C_{22}(\theta)$ discussed above.
Thus the zero-lag noise variance can be written as $\sigma^2_0=\sigma^2_{0ran}+\sigma^2_{0corr}$, 
where
\begin{equation}\label{w1}
\sigma^2_{0ran}={\sigma^2_{\epsilon}\over 2 n_g}(2\pi)^2\int d\vec k
\hbox{ }|\tilde W(\vec k)|^2,
\end{equation}
and 
\begin{equation}\label{w1}
\sigma^2_{0corr}=(2\pi)^4\int d\vec k\hbox{ }|\tilde W(\vec k)|^2
{1\over 2}\bigg [\tilde C_{11}(\vec k)+\tilde C_{22}(\vec k)\bigg ],
\end{equation}
where the factor $1/2$ is from the integration of $c_1^2(k)=\cos^2(2\phi)$ 
[and $c_2^2(k)=\sin^2(2\phi)$] over $\phi$.

Considering Gaussian smoothings with 
\begin{equation}\label{w1}
W(\theta)={1\over \pi\theta_G^2}\exp\bigg (-{\theta^2\over \theta_G^2}\bigg ),
\end{equation}
where $\theta_G$ is the angular smoothing scale, we have 
\begin{equation}\label{w1}
\sigma^2_{0ran}={\sigma^2_{\epsilon}\over 2}{1\over 2\pi\theta_G^2 n_g},
\end{equation}
and \begin{equation}\label{w1}
\sigma^2_{0corr}={1\over 2 \pi}\int d\vec \theta \hbox{ }{1\over 2}\bigg [C_{11}(\theta )
+C_{22}(\theta)\bigg ] {1\over \theta_G^2}\exp\bigg (-{\theta^2\over 2\theta_G^2}\bigg ).
\end{equation}

In the following analyses, we use eq. (4), (6) and (7) to calculate $C_{11}(\theta)+C_{22}(\theta)$,
and further $\sigma^2_{0corr}$ from eq. (23). 

The angular correlations $C_{ij}(\theta)$ depend sensitively on the redshift distribution of 
background galaxies. For galaxies distributed in a narrow range around a relatively low redshift, 
a large fraction of them are physically close to each other, resulting large $C_{ij}(\theta)$. 
We adopt the following functional form to describe the distribution of background galaxies 
\begin{equation}\label{w1}
p(z)={\beta \over \Gamma [(1+\alpha)/\beta]}\bigg ({z\over z_s}\bigg )^{\alpha}\exp\bigg [-\bigg ({z\over z_s}\bigg )^{\beta}\bigg ],
\end{equation}
where $\alpha, \beta$ and $z_s$ are parameters that can be determined from survey conditions.
We take $\alpha=2$ and $z_s=0.7$. To see the effect of the width of the distribution, 
we vary the $\beta$ value with $\beta=1, 1.5, 3, $ and $6$. The larger the $\beta$ value is,
the narrower the distribution is, as seen in Figure 1. The corresponding median redshifts
for the four distributions are $z_{med}\approx 1.87$, $0.99$, $0.62$ and $0.55$ for 
$\beta=1, 1.5, 3, $ and $6$, respectively.

In Figure 2, we show the results of $\sigma^2_{0corr}$. For the intrinsic alignment,
we take $A=0.57$, the value from SDSS, in eq. (4) (e.g., Heymans et al. 2006). 
The solid, dotted, dashed, 
and dash-dotted lines correspond to $\beta=6, 3, 1.5$ and $1$, respectively.
For comparison, we also plot $\sigma^2_{0ran}$ (dash-dot-dot-dotted line)
with $\sigma_{\epsilon}=0.4$ and $n_g=30\hbox{ arcmin}^{-2}$. 
We see that the result with $\beta=6$ and $z_{med}\sim 0.55$ is an order of 
magnitude larger than that with $\beta=1$ and $z_{med}\sim 1.87$,
demonstrating clearly the sensitive dependence of $\sigma_{0corr}$ on the
redshift distribution of background galaxies. Therefore for tomographic analyses of
weak lensing effects with source galaxies distributed in narrow redshift bins, 
the effects of intrinsic alignments 
can be significant. The angular dependence of $\sigma^2_{0corr}$ is shallower 
than $\sigma^2_{0ran}$, and the ratio of $\sigma^2_{0corr}/\sigma^2_{0ran}$ 
increases with the increase of smoothing scales. 

In Table 1, we list $\sigma^2_{0corr}/\sigma^2_{0ran}$ for various cases.
With the upper limit $A=1.29$ from SDSS, the ratio 
can reach as high as about $\sigma^2_{0corr}/\sigma^2_{0ran}\sim 20\%$ for $\beta=6$ 
at $\theta_G=2 \hbox{ arcmin}$.
Notice that $\sigma^2_{0ran}$ and $\sigma^2_{0corr}$ depend differently on the 
distribution of source galaxies. While $\sigma^2_{0corr}$ depends mainly 
on the form of the redshift distribution, $\sigma^2_{0ran}\propto n_g^{-1}$. 
Thus for surveys with higher surface number density of source galaxies 
than what we consider here, the ratio $\sigma^2_{0corr}/\sigma^2_{0ran}$
can increase considerably. Results expected for some surveys are presented in 
Table 2. The survey parameters for COSMOS are taken from Massey et al. (2007).
For SNAP, we adopt the parameters used in Semboloni et al. (2007).
For deep surveys with large $n_g$, tomographic analyses with source galaxies distributed 
in narrow redshift ranges become possible. For example, with total $n_g\sim 100\hbox { arcmin}^{-2}$
as expected from surveys similar to SNAP, the background galaxies can be divided into three 
bins each with $n_g\sim 30\hbox { arcmin}^{-2}$.
The effect of intrinsic alignments can be significantly stronger within each bin than that
in total. If we regard the narrow redshift distribution with $\beta=6$ as one of the 
bins, it is seen from Table 1 that the respective values of $\sigma^2_{0corr}/\sigma^2_{0ran}$ for
$\theta_G=1 \hbox{ and } 2\hbox{ arcmin}$ are about $5\%$ and $10\%$ with $A=0.57$,
in comparison with $3.3\%$ and $5\%$ expected for the full sample of galaxies from SNAP
as seen in Table 2.

In next section, we show that the number of false peaks in lensing $\kappa$-maps is
very sensitive to the ratio of $\sigma^2_{0corr}/\sigma^2_{0ran}$.
Even a relatively low value of $\sigma^2_{0corr}$ can result a considerable increase of the
number of false peaks, and therefore reduce the efficiency of cluster detections significantly.

\section{Statistics of false peaks in $\kappa$-maps resulting from intrinsic ellipticities}

Weak lensing cluster detections associate high peaks in $\kappa$-maps reconstructed from shear
measurements with clusters of galaxies. Intrinsic ellipticities of background galaxies
can produce false peaks, and therefore affect the efficiency of 
cluster detections. It is thus important to understand the statistics of false peaks 
thoroughly in order to extract reliable cluster samples from weak lensing surveys.
van Waerbeke (2000) studies the number of false peaks assuming no intrinsic alignments 
for source galaxies. In this case, the smoothed quantity $N(\vec \theta)$ 
defined in eq. (14) is approximately a Gaussian random field because of the
central limit theorem. Including the correlations of the intrinsic ellipticities, the 
statistics of the noise field $N(\vec \theta)$ can be complicated. 
Relating the intrinsic ellipticities linearly with the tidal field predicts Gaussian statistics. 
Assuming that they are associated with galaxy spins
gives rise to non-Gaussian statistics for the intrinsic ellipticities. On the other hand,
$N(\vec \theta)$ is related to the sum of the intrinsic ellipticities of background galaxies
in the smoothing window. Since the intrinsic alignments are relatively weak 
(on the order of $10^{-6}$ to $10^{-5}$ at $\theta_G\sim 1\hbox{ arcmin}$), 
according to the central limit theorem we do not expect a highly non-Gaussian field
for $N(\vec \theta)$ if the number of galaxies within the smoothing window is large enough.
In our following analyses, we assume the Gaussianity for $N(\vec \theta)$.  
Detailed studies on its statistics will be carried out in our future investigations. 

For a two dimensional Gaussian random field $N$, the differential number density of peaks can be 
written explicitly in the following form (Bond \& Efstathiou 1987; van Waerbeke 2000)
\begin{equation}\label{w1}
n_{peak}(\nu)={1\over 2\pi \theta_*^2}\exp(-\nu^2/2){G({\gamma_p}, {\gamma_p}\nu)\over \sqrt{2\pi}},
\end{equation}
where $\nu=N/\sigma_0$ is the significance of a peak with $N$ being the value of
the considered quantity at the peak position, $b=\sqrt{2(1-{\gamma_p}^2)}$, and
\begin{eqnarray}\label{w13}
G({\gamma_p}, \hat x)&=&{1\over 2}\bigg ({\hat x}^2+{b^2\over 2}-1\bigg )\hbox{ erfc}\bigg (-{\hat x\over b}
\bigg )+{{\hat x}b\over 2\sqrt{\pi}}\exp(-{\hat x}^2/b^2) \nonumber\\
& &  +{1\over 2(1+b^2)^{1/2}}\exp[-{\hat x}^2/(1+b^2)]\hbox{ erfc}\bigg (-{{\hat x}\over b\sqrt{1+b^2}}\bigg ).
\end{eqnarray}
It is seen that $n_{peak}(\nu)$ is fully characterized by
${\gamma_p}$ and $\theta_{*}$, which are respectively defined as
\begin{equation}\label{w1}
\gamma_p={\sigma_1^2\over \sigma_0 \sigma_2} \quad \hbox{ and} \quad \theta_{*}=\sqrt{2}{\sigma_1\over \sigma_2},
\end{equation}
where 
\begin{equation}\label{w1}
\sigma_n^2=\int {d\vec k}\hbox{ }k^{2n}<|N(k)|^2>.
\end{equation}

Considering the noise field $N(\theta)$ defined in eq. (14), 
with Gaussian smoothings we have $\gamma_p=\sqrt{2}/2$ 
and $\theta_{*}=\theta_G/\sqrt{2}$ in the case without intrinsic alignments (van Waerbeke 2000). 
Thus the average cumulative number density of peaks 
$N_{peak}(\nu_{ran})=\int_{\nu_{ran}}n_{peak}(\nu')d\nu'$ is 
independent of $\sigma_{0ran}$ and scales with the smoothing angle as $\theta_G^{-2}$. 
In this case, given a survey area and a smoothing angle $\theta_G$, the average 
number of false peaks in terms of the significance $\nu_{ran}$ is fixed 
regardless of the specific value of $\sigma_{0ran}$. Note that 
the existence of false peaks is the result of chance alignments of background 
galaxies. Given $\nu_{ran}=N/\sigma_{0ran}$ for a false peak, its
strength $N$ is proportional to $\sigma_{0ran}$ which is in turn determined 
by $n_g$ and $\sigma_{\epsilon}$. Thus the number of false peaks measured by their
strength $N$ depends on $n_g$ and $\sigma_{\epsilon}$. Smaller $n_g$
or larger $\sigma_{\epsilon}$ gives rise to larger $\sigma_{0ran}$ and
leads to higher probabilities in forming false peaks with large $N$ by 
chance alignments. 

Including the intrinsic alignments, both $\gamma_p$ and $\theta_{*}$,
and thus the number density of false peaks in terms of the true significance $\nu$, 
depend on the correlation level. It is noted from Figure 2 that the level of intrinsic alignments
from current observations is low comparing with $\sigma_{0ran}$. Therefore 
$\gamma_p$, $\theta_{*}$, and $n_{peak}(\nu)$ change only slightly with respect to the case without
intrinsic alignments. 


The number of false peaks discussed above is given
in terms of the true significance $\nu=N/\sigma_0$, i.e.,
the peak height is measured relative to the full noise variance 
$\sigma_0=\sqrt{\sigma^2_{0ran}+\sigma^2_{0corr}}$ including 
$\sigma_{0corr}$.  Observationally however, it is difficult to obtain
$\sigma_{0corr}$,  and thus usually only $\sigma_{0ran}$ is estimated and 
used in measuring the significance of a peak. The true significance of the peak
corresponding to the observed significance $\nu_{ran}=N/\sigma_{0ran}$
is then $\nu=\nu_{ran}/(1+\sigma^2_{0corr}/\sigma^2_{0ran})^{1/2}$.
Given a threshold on $\nu_{ran}$, the average number of peaks is 
$N_{peak}(\nu_{ran})\propto \int_{\nu} n_{peak}(\nu^{'}) d\nu^{'}$.  
Because $\nu<\nu_{ran}$ for non-zero $\sigma_{0corr}$, 
$N_{peak}$ increases with the increase of $\sigma_{0corr}$.

In Figure 3, we show $N_{peak}-N^{ran}_{peak}$, the number of false peaks
resulting from intrinsic alignments, with respect to the detection threshold $\nu_{ran}$,
where $N_{peak}$ and $N^{ran}_{peak}$ are the cumulative numbers of false peaks 
in $1 \hbox{ deg}^{2}$ with and without intrinsic alignments. 
For comparison, we also plot $N^{ran}_{peak}$ (thick lines) in each panel. 
The left and right panels are respectively for $\theta_G=1\hbox{ arcmin}$ and
$\theta_G=2\hbox{ arcmin}$. The upper panels are for $A=0.57$, and
the lower panels are for $A=1.29$. The thin solid, dotted, dashed, and dash-dotted lines
in each panel are for $\beta=6$, $3$, $1.5$ and $1$, respectively.
It is seen that $N_{peak}-N^{ran}_{peak}$ can be comparable to $N^{ran}_{peak}$.
For $\theta_G=1\hbox { arcmin}$, $N^{ran}_{peak}\sim 2$, $0.3$, $0.04$, and $0.004$
at $\nu_{ran}=3.5$, $4$, $4.5$ and $5$, respectively. The corresponding 
$N_{peak}-N^{ran}_{peak}$ are $0.25$, $0.06$, $0.013$, and $0.0015$ for
$\beta=1.5$ and $A=1.29$. For larger $\beta$, the numbers are larger and
$N_{peak}-N^{ran}_{peak}>N^{ran}_{peak}$ when $\nu_{ran}>3.7$ and $4.5$ for
$\beta=6$ and $3$, respectively. For $\theta_G=2\hbox{ arcmin}$, 
$N_{peak}-N^{ran}_{peak}>N^{ran}_{peak}$ when $\nu_{ran}>3$, $3.8$ and
$5.3$ for $\beta=6$, $3$ and $1.5$, respectively. Therefore the
existence of intrinsic alignments can result significant number of 
extra false peaks in lensing convergence maps. 

In Figure 4, we show the dependence of the ratio $r_{peak}=N_{peak}/N^{ran}_{peak}$
on the level of intrinsic alignments represented by the amplitude $A$ for 
$\theta_G=1\hbox{ arcmin}$.
The $\beta$ value in each panel is written out explicitly. The 
solid, dotted, dashed and dash-dotted lines are respectively for $\nu_{ran}=5$, $4.5$,
$4$ and $3.5$. For $\beta=1.5$ and $A=1.29$, we have $r_{peak}\sim 1.36$, 
$1.27$, $1.2$ and $1.14$ for $\nu_{ran}=5$, $4.5$, $4$ and $3.5$, respectively.
With larger smoothing scales, the relative effect of intrinsic alignments is higher.
For $\theta_G=2\hbox{ arcmin}$, the corresponding
ratios change to $1.6$, $1.4$, $1.3$ and $1.2$. For $\beta=6$, the ratios for $\nu_{ran}=5$ reach
as high as $3.8$ and $7.6$ for $\theta_G=1$ and $2\hbox{ arcmin}$, respectively.

From eqs. (25) and (26), it can be shown that $N_{peak}$ depends largely  
on the detection threshold with $N_{peak}\propto \nu\exp(-\nu^2/2)$ at $\nu>3$ 
(e.g., van Waerbeke 2000). Given a detection threshold on $\nu_{ran}$, the 
corresponding threshold for the true significance is
$\nu=\nu_{ran}/(1+\sigma^2_{0corr}/\sigma^2_{0ran})^{1/2}$, which decreases with 
the increase of $\sigma^2_{0corr}/\sigma^2_{0ran}$.
Thus $r_{peak}$ is largely determined by the ratio $\sigma^2_{0corr}/\sigma^2_{0ran}$. 
In Figure 5, we show $r_{peak}$ with
respect to $\sigma^2_{0corr}/\sigma^2_{0ran}$ for $\theta_G=1$ (upper panel) and $2\hbox{ arcmin}$
(lower panel). The four sets of lines from top to bottom in each panel correspond 
respectively to the threshold $\nu_{ran}=5$, $4.5$, $4$ and $3.5$. Note that each set contains
four lines with $\beta=6$, $3$, $1.5$, and $1$, respectively. 
With $\sigma^2_{0corr}/\sigma^2_{0ran}\sim 5\%$, the values of $r_{peak}$ are about 
$1.7$, $1.55$, $1.4$ and $1.3$ for $\nu_{ran}=5$, $4.5$, $4$ and $3.5$. For
$\sigma^2_{0corr}/\sigma^2_{0ran}\sim 10\%$, the corresponding $r_{peak}$ are 
$2.9$, $2.3$, $1.9$ and $1.6$. A specific value of $\sigma^2_{0corr}/\sigma^2_{0ran}$
depends on the strength of the intrinsic alignment, the surface number density and 
the redshift distribution of source galaxies, and $\sigma_{\epsilon}$.
The dotted vertical lines from left to right in each panel
show the corresponding values of $\sigma^2_{0corr}/\sigma^2_{0ran}$ for
$\beta=1$, $1.5$, $3$ and $6$, where we take $A=1.29$, $\sigma_{\epsilon}=0.4$,
and $n_g=30\hbox{ arcmin}^{-2}$. It should be noted that $n_g$ usually varies with the 
redshift distribution of source galaxies. Surveys that can reach high redshifts typically
have large $n_g$. Thus our estimates on $\sigma^2_{0corr}/\sigma^2_{0ran}$
with fixed $n_g=30\hbox{ arcmin}^{-2}$ may overestimate the ratio for $\beta=6$. 
On the other hand, for deep surveys with large $n_g$ (e.g., $n_g\sim 100$ for SNAP, and 
$n_g\sim 300$ for SNAP Deep), we can divide the source galaxies into different bins 
with $n_g\sim 30 \hbox{ arcmin}^{-2}$ in each bin. In this case, 
the narrow distribution with $\beta=6$ can be one of these bins, and our
above estimate on $\sigma^2_{0corr}/\sigma^2_{0ran}$ 
with $n_g=30\hbox{ arcmin}^{-2}$ can be a representative value for galaxies within
the bin.  
  
For weak lensing cluster surveys, the efficiency 
measures how efficient we can find true clusters from lensing maps. Assuming the
NFW profile (Navarro, Frenk and White 1996) for the mass distribution of clusters 
of galaxies, it is found
that the number of peaks resulting from true clusters in lensing $\kappa$-maps 
is about $6\hbox{ deg}^{-2}$, and $4\hbox{ deg}^{-2}$ for significance larger than $3.5$ and $4$, 
respectively, where $\sigma_{\epsilon}=0.4$, $n_g=30\hbox{ arcmin}^{-2}$ and 
$\theta_G=1\hbox{ arcmin}$ are used (Hamana et al. 2004). As we show previously, 
the corresponding $N^{ran}_{peak}$ are about $2$ and $0.3$. Then a simple estimate
gives the efficiency about $75\%$ and $93\%$ for the two detection thresholds if there are
no intrinsic alignments. With $\sigma^2_{0corr}/\sigma^2_{0ran}\sim 10\%$, 
the corresponding efficiencies drop to $65\%$ and $87\%$. Note that we only
consider the contamination from false peaks when estimating the above efficiencies. The
existence of intrinsic ellipticities and alignments not only results false peaks
but also affects the heights of true peaks, which can further decrease the efficiency 
of weak lensing cluster detections considerably (e.g., Hamana et al. 2004). 

Because weak lensing effects arise from the gravitational influence of the matter distribution,
it is expected that dark clumps without luminous counterparts can be discovered
from lensing observations. The existence of massive dark clumps would question the current theory of
structure formation seriously. There have been such candidates reported in literature.
Erben et al. (2000) present CFHT weak lensing results around the galaxy cluster Abell 1942. 
They find a high peak [$\sim 5\sigma$ in the aperture-mass measurement (Schneider 1996)] 
without associated galaxy overdensities at the location about $7'$ south 
of the main cluster. Faint X-ray emissions from the nearby region of the peak
are detected by ROSAT, but they may not be related to the lensing peak signal. 
With HST lensing observations in this field, 
Linden et al. (2006) also find a peak at a place consistent with that given by 
Erben et al. (2000), but with a much lower significance $\sim 2.9\sigma$. 
Further, they divide the source galaxies into three magnitude bins and perform
lensing analyses for each of them. For the bright bin, 
which contains most of the source galaxies used in Erben et al. (2000), they find 
a $1.9\sigma$ peak with a smoothing scale $120''$ . For the faint bin,
a $3.3\sigma$ peak is detected. There is no lensing detection
from the medium bin, which is unexpected if there is a foreground dark clump. 
A spatial concentration of galaxies in the medium bin is observed, which could act as the 
lens for galaxies in the faint bin but not for those in the bright bin. 
The lack of lensing detections in the medium bin and the low significance of the peak 
from HST observations raise questions on the lensing origin of the peak. It is likely that
the peak is a statistical fluke (Linden et al. 2006). On the other hand,
Erben et al. (2000) estimate the probability that their detected peak is a false one from
chance alignments of background galaxies.
In order to apply the results given by van Waerbeke (2000), they perform a Gaussian smoothing 
with $\theta_G\approx 0.5'$ to the $\kappa$ field, and find that the considered peak 
has a height of $\nu_{ran}\sim 4.5\sigma$. The probability to have such a high peak from
chance alignments is very low (Erben et al. 2000). 

Our analyses show that the existence of intrinsic alignments can increase
the chance for the appearance of false peaks in a given
area depending on the ratio of $\sigma^2_{0corr}/\sigma^2_{0ran}$. 
In Table 3, we list the probabilities that the detected 'dark clumps' are false peaks 
for different observations. The probability is calculated from the Poisson statistics 
with 
\begin{equation}\label{w1}
p_n={e^{-N_{peak}}N_{peak}^{n}\over n!},
\end{equation}
where $p_n$ is the probability to have $n$ false peaks in a field, and $N_{peak}$ is the 
average number of false peaks expected in the field. 
%
For the observation of Erben et al. (2000), we estimate $\sigma^2_{0corr}/\sigma^2_{0ran}\sim 1.3\%$
(with $A=1.29$) for $\theta_G=0.5\hbox{ arcmin}$. With this level of intrinsic alignment,
the probability to find one $\nu_{ran}=4.5$ false peak in the field of $14'\times 14'$ increases 
only slightly from $0.9\%$ to $1\%$. 
On the other hand, with the noise level comparable to that of Erben et al. (2000), 
the average surface number density of peaks with $\nu_{ran}\ge 4.5$
resulting from the lensing effects of true mass concentrations is 
about $3\hbox{ deg}^{-2}$ for $\theta_G=0.5\hbox{ arcmin}$ (Hamana et al. 2004). 
Then the average number of true peaks in the field of $14'\times 14'$ is about $0.15$, and
the probability to have one true peak in this field is $\sim 13\%$.
Thus the $\nu_{ran}=4.5$ peak found by Erben et al. 
(2000) is much more likely to be associated with a true mass clump
than being a false peak.  However the analyses on HST data by Linden et al. (2006) give 
$\nu_{ran}\sim 3$ for the peak. The average number of 
false peaks expected in the field of $14'\times 14'$ with $\nu_{ran}\ge 3$ is about $1.7$ with
$\sigma^2_{0corr}=0$. For source galaxies in Linden et al. (2006), we estimate 
$\sigma^2_{0corr}/\sigma^2_{0ran}\sim 5\%$ with $A=1.29$.
Then the average number of false peaks increases to $\sim 2$, and the 
corresponding probability $p(corr)\sim 27\%$.
In this case, it is quite possible that the observed peak is a false one.
The reason for the difference between the peak heights from 
CFHT and HST is unclear (Linden et al. 2006), and so is the conclusion on the origin of the peak. 
It is likely that the peak is associated with a small mass clump
and its height is enhanced by the chance alignment of background galaxies (Linden et al. 2006). 

Massey et al. (2007) present the COSMOS $2\hbox{ deg}^2$ lensing analysis. They 
notice the existence of two high peaks without luminous counterparts 
near the main cluster. The $\kappa$ field in Massey et al. (2007) is reconstructed using 
the wavelet method and its noise properties are complicated 
(e.g., Starck, Pires, \& Refregier 2006). The 
significances of the peaks are not clearly given in Massey et al. (2007). 
Thus in Table 3, we include the probabilities for different significances.
When calculating $p(corr)$, we use $\sigma^2_{0corr}/\sigma^2_{0ran}\sim 5\%$ estimated 
for the redshift distribution of source galaxies with $\alpha=2$, $\beta=1.5$ and $z_s=0.8$. 
It is seen that with the intrinsic alignments, the probability that the two peaks 
are false ones is tripled for $\nu_{ran}>5$.  
It is noted that our results are for Gaussian smoothings, and therefore 
cannot be used directly to discuss how likely the peaks 
found by Massey et al. (2007) are false ones. 
On the other hand, the COSMOS data can be readily analyzed with the 
method of Kaiser \& Squires (1993) with Gaussian smoothings. Then
our studies presented here can be directly applicable.

\section{Constraints on intrinsic alignments from CFHTLS Deep survey} 

Because of its sensitive dependence on intrinsic
alignments, the number of false peaks can be used to probe the strength of
the intrinsic alignments of source galaxies
if one can observationally distinguish false and true peaks.
In this section, we analyze the constraints on $\sigma^2_{0corr}$ from the
results of the CFHTLS Deep survey (Gavazzi \& Soucail 2007). 

The CFHTLS Deep survey shares the same data with the Supernova Legacy Survey (SNLS).
It contains four independent fields and includes data from five bands $(g^{\prime},
r^{\prime},i^{\prime},z^{\prime},u^{*})$. The shear measurements are done using 
the $i^{\prime}$ band images with magnitude in the range $22<i^{\prime}<26$. The seeing is
$\sim 0.9^{\prime\prime}$. The total working area for weak lensing analysis
is $3.61\hbox{ deg}^2$. The photometric redshift
is estimated for each source galaxy with the multi-band observational data. The 
redshift distribution for a subsample of source galaxies with reliable photo-z
measurements is presented in Figure 3 of Gavazzi \& Soucail (2007),
which will be used in our following analysis. The convergence $\kappa$-map is constructed
from the inferred shear $\gamma$ with the technique developed by Kaiser and Squires (1993). 
The shear and consequently the $\kappa$ fields are smoothed with a Gaussian window function
with $\theta_G=1\hbox{ arcmin}$. The variance of noise in the smoothed $\kappa$ field from
randomly oriented background galaxies is estimated to be $\sigma_{0ran}=0.0196, 0.0225, 
0.0202$, and $0.0221$ for the four fields, respectively (Gavazzi \& Soucail 2007).
The signal-to-noise ratio $\nu$ is defined as $\nu_{ran}=\kappa/\sigma_{0ran}$. From the
$\kappa$-maps of the four fields, Gavazzi and Soucail (2007) detect $\sim 46$ peaks
with $\nu_{ran}>3$, $14$ peaks with $\nu_{ran}>3.5$ and $5$ peaks with $\nu_{ran}>4$. 
Detailed studies are done for the $14$ peaks with $\nu_{ran}>3.5$. 
With the help of photometric redshift measurements, X-ray observations and the 
lensing tomographic analysis, they claim that there are $9$ secure cluster detections 
among the $14$ peaks. The rest $5$ are likely false peaks. 
In our study here, we regard these $5$ peaks as
false ones resulting from the intrinsic ellipticities of background galaxies, 
and constrain the level of intrinsic alignments based on our analysis
presented in the previous section. It is worth mentioning that the statistics
based on only $5$ peaks is poor, and therefore our analysis mainly aims at demonstrating
the feasibility in extracting the information of the intrinsic alignments of background
galaxies from the number of false peaks. Also some of the five peaks may result from 
dark clumps without luminous counterparts. Larger weak lensing cluster surveys 
with more reliable tomographic analyses will provide statistically meaningful results on 
the intrinsic alignments.

We model the redshift distribution shown in Gavazzi \& Soucail (2007) as
\begin{equation}\label{w12}
p(z)\propto \bigg ( {z\over z_s}\bigg )^2 \exp \bigg [-\bigg ({z\over z_s}\bigg )^2\bigg ]
+0.07\exp\bigg [-{(z-2.8)^2\over 0.6^2}\bigg ],
\end{equation}
where $z_s$ is taken to be $0.8$. The second term is added to describe the low
bump at $z\sim 3$ seen in the redshift distribution of the source galaxies of CFHTLS Deep
(Gavazzi \& Soucail 2007). This term does not affect much our results because of its
low amplitude. We calculate $\sigma_{0corr}$ from eq.(23) with $\theta_G=1\hbox{ arcmin}$,
where $C_{11}$ and $C_{22}$ are computed from eqs. (4)-(7).

In Figure 6, we show the dependence on $\sigma^2_{0corr}/\sigma^2_{0ran}$ of the expected
number of false peaks detected in $3.61\hbox{ deg}^2$. 
The solid line is for the average number of false peaks.
The upper and lower dashed lines show the $-1\sigma$ and $-2\sigma$ Poisson deviations from the 
mean, i.e., $N_{peak}-\sqrt{N_{peak}}$ and $N_{peak}-2\sqrt{N_{peak}}$, respectively.
The horizontal dash-dotted line is located at $N_{peak}=5$. The vertical dotted lines
indicate the values of $\sigma^2_{0corr}/\sigma^2_{0ran}$ for the source galaxies of
CFHTLS Deep with $A=0.57$ and $A=1.29$, from left to right, respectively. 
We see that $1\sigma$ and $2\sigma$ constraints give $\sigma^2_{0corr}/\sigma^2_{0ran}<4\%$
and $<14\%$. With $\sigma^2_{0ran}=0.0004$, we have $\sigma^2_{0corr}<1.6\times 10^{-5}$
and $\sigma^2_{0corr}< 5.6\times 10^{-5}$ for $1\sigma$ and $2\sigma$ constraints. 
From eq. (23), it is seen that $2\sigma^2_{0corr}$ corresponds
to the intrinsic alignment $C_{11}+C_{22}$ smoothed over the angular
scale $\theta_G$. Thus we have the $1\sigma$ and $2\sigma$ constraints on 
$C_{11}+C_{22}$ for $\theta_G=1\hbox{ arcmin}$ to be $<3.2\times 10^{-5}$ and $<1.1\times 10^{-4}$.
The corresponding limits on the amplitude of the intrinsic alignments
$A$ are $A<2.9$ and $A<10$. The results are fully consistent with that from SDSS observations
with $A<1.29$.


Schirmer et al. (2007) analyze a total of $\sim 20\hbox{ deg}^{2}$ data collected
from different observations with different observational depth. 
They present a sample of shear-selected clusters containing total
$158$ candidates identified by two types of statistics.  
Using only $S$-statistics which is similar to the aperture mass statistics but with 
different filtering functions, they find $91$ peaks with significance being higher than $4$. 
Among them, there are $48$ dark ones without obvious optical counterparts. It is found that
the fraction of dark peaks is relatively high in shallow surveys with 
low surface number density of source galaxies. This indicates that a significant 
number of dark peaks could be false ones resulting from intrinsic ellipticities. 
Since their filtering functions are complicated with different filtering scales, 
we cannot do quantitative analyses on the constraints on intrinsic alignments with 
these dark peaks. However, we may give some rough estimates. 
In Schirmer et al. (2007), the filtering scales used in peak identifications
range from $1.6^{'}$ to $19.8^{'}$ with most of them being larger than $2^{'}$. 
Comparing the functional form of the Gaussian smoothing with the filtering functions
used in Schirmer et al. (2007), their filtering should have the effects corresponding to
Gaussian smoothings with $\theta_G\ge 1^{'}$.
Thus we use the results of $\theta_G=1^{'}$ for a conservative discussion.
Without intrinsic alignments, our results show that
the average number of false peaks in $20\hbox{ deg}^{2}$ with $\nu_{ran}\ge 4$ is
about $6$. The existence of intrinsic alignments enhances the average number of false peaks.
If the $48$ dark peaks are all false ones and the number is $+1\sigma$ from the average number
of false peaks, we need the average number to be $N_{peak}\sim 42$. Then we have to have 
$\sigma^2_{0corr}/\sigma^2_{0ran}\sim 35\%$ to get such a high number of false peaks.
For the redshift distribution of source galaxies, Schirmer et al. (2007) 
give $\alpha=2$, $\beta=1.5$ and $z_s=0.4$ for shallow surveys. 
The surface number density is $n_g\sim 12\hbox{ arcmin}^{-2}$, and $\sigma_{\epsilon}\sim 0.48$
(Schirmer et al. 2007). We then estimate $\sigma^2_{0ran}\sim 0.0015$.
Thus the ratio $\sigma^2_{0corr}/\sigma^2_{0ran}\sim 35\%$ requires the parameter $A$ to be $A\sim 32$,
which is much higher than the constraint $A\le 1.29$ from SDSS.  
Therefore it is very unlikely that the $48$ dark peaks are all false ones from 
intrinsic ellipticities of background galaxies. 
As we discussed previously and also in Schirmer et al. (2007),
the joint effects of small mass clumps and the intrinsic ellipticities could contribute
significantly to the number of dark peaks with high significance.
It should be pointed out that
the functional form and the scale of the filtering function adopted by Schirmer et al. (2007) 
are optimized to detect clusters with NFW density profiles. With Gaussian smoothings,
the number of peaks and their properties may change quantitatively. It is therefore
desirable to analyze the observations with Gaussian smoothings so that we can perform detailed
analyses on the statistics of false peaks. On the other hand, it is also
worthwhile investigating the noise properties and the associated statistics of false peaks 
under different smoothing schemes. As the catalog of Schirmer et al. (2007) is the
largest one so far, from many aspects careful observational and theoretical 
studies on these dark peaks are highly valuable. 

Future surveys with larger areas will result many more peaks. If a large number of false peaks
from intrinsic ellipticities can be securely identified, we can put tight
constraints on the level of intrinsic alignments. Considering Poisson fluctuations, we 
can estimate, as follows, how well the quantity $x=\sigma^2_{0corr}/\sigma^2_{0ran}$ can be constrained
from $N$ false peaks with $\nu_{ran}\ge \nu_0$. With the average cumulative number of peaks
$N_{peak}\propto \nu \exp(-\nu^2/2)$ for $\nu\ge 3$, where $\nu$ is the true significance 
(e.g., van Waerbeke 2000), we have, for the
central value of $x$, denoted by $x_c$,
\begin{equation}
N_{ran}{(\nu_0/\sqrt{1+x_c})\exp[ -\nu_0^2/ 2/(1+x_c)]\over 
\nu_0\exp[ -\nu_0^2/ 2]}={N_{ran}\over \sqrt{1+x_c}}\exp\bigg [ {\nu_0^2 x_c\over 2(1+x_c)}\bigg ]=N,
\end{equation}
where $N_{ran}$ is the average number of false peaks expected in the field 
without considering intrinsic alignments.
The $\pm 1\sigma$ constraints on $x$ can then be obtained by
\begin{equation}
{N_{ran}\over \sqrt{1+x_c + \delta x}}\exp\bigg [ {\nu_0^2 (x_c + \delta x)\over 2(1+x_c + \delta x)}\bigg ]-
\sqrt{{N_{ran}\over \sqrt{1+x_c + \delta x}}\exp\bigg [ {\nu_0^2 (x_c + \delta x)\over 2(1+x_c + \delta x)}\bigg ]}=N,
\end{equation}
and
\begin{equation}
{N_{ran}\over \sqrt{1+x_c - \delta x}}\exp\bigg [ {\nu_0^2 (x_c - \delta x)\over 2(1+x_c - \delta x)}\bigg ]+
\sqrt{{N_{ran}\over \sqrt{1+x_c - \delta x}}\exp\bigg [ {\nu_0^2 (x_c - \delta x)\over 2(1+x_c - \delta x)}\bigg ]}=N.
\end{equation}
With $\delta x<<1$, we have
\begin{equation}
\delta x\approx {1\over \nu_0^2/2-1/2}{\sqrt{N}\over N}{1\over 1+0.5/\sqrt{N}}.
\end{equation}
For instance, with $N=50$ and $\nu_0=4$, we have $\delta x\sim 1.8\%$, i.e.,  
the quantity $x$ can be constrained to the level of $x_c\pm 1.8\%$ ($1\sigma$).
From $\delta x$ to $\delta (\sigma^2_{0corr})$, it depends on $\sigma^2_{0ran}$, and thus 
on $n_g$ and $\sigma_{\epsilon}$. Further from the constraint on $\sigma^2_{0corr}$ to 
the constraint on $A$, we need the redshift distribution of background galaxies.
With $n_g=30\hbox{ arcmin}^{-2}$, $\sigma_{\epsilon}=0.4$, $\alpha=2$, $\beta=1.5$, 
and $z_s=0.7$, we have $\delta (\sigma^2_{0corr})\sim 7\times 10^{-6}$, and
$\delta (A)\sim 0.9$. From the relation between $2\sigma^2_{0corr}$ and 
the intrinsic alignment $C_{11}+C_{22}$ smoothed over the angular
scale $\theta_G$, the limit $\delta (\sigma^2_{0corr})\sim 7\times 10^{-6}$ leads to
$\delta (C_{11}+C_{22})\sim 1.4\times 10^{-5}$ over $\theta_G=1^{'}$.


Intrinsic alignments of galaxies have been estimated from nearby surveys SuperCOSMOS
(Brown et al. 2002) and SDSS (Mandelbaum et al. 2006) assuming negligible lensing effects.
Extrapolating their results to redshifts $z\sim 1$ appropriate for most lensing surveys
suffers many uncertainties (Brown et al. 2002; Mandelbaum et al. 2006). Heymans et al. (2004)
estimated the intrinsic alignments in the COMBO-17 survey ($z\sim 0.6$) from close
pairs of background galaxies. Their error bar at $\sim 1\hbox{ arcmin}$ is about 
$a\hbox{ } few \times 10^{-4}$. Our above analysis based on only $5$ false peaks already gives rise 
to a tighter constraint on the order of $\delta (C_{11}+C_{22})\sim 4\times 10^{-5}$, 
demonstrating the great potential of our proposed method.  

\section{Discussion}

In this paper, we investigate the effect of the intrinsic alignments of background galaxies
on weak lensing detections of mass concentrations. 
Focusing on the convergence $\kappa$-maps, 
we analyze the number of false peaks due to the
intrinsic ellipticities of background galaxies taking into account their intrinsic alignments.  
Under the assumption of Gaussianity for the noise field, the number of false peaks 
in $\kappa$ maps depends on two characteristic parameters $\gamma_p$ and $\theta_*$,
which are in turn determined by the two-point correlations of the field. Without 
intrinsic alignments, $\gamma_p=\sqrt{2}/2$ and $\theta_*=(\sqrt{2}/2)\theta_G$ for 
a Gaussian window. Thus the number of false peaks
in terms of $\nu_{ran}=N /\sigma_{0ran}$ does not depend on $\sigma_{0ran}$ 
(Note that given $\nu_{ran}$, the strength of a peak $N$ depends on $\sigma_{0ran}$).
With the intrinsic alignments, however, both $\gamma_p$ and $\theta_*$ change with
$\sigma_{0corr}$. More importantly, the full noise variance
$\sigma^2_{0}=\sigma^2_{0ran}+\sigma^2_{0corr}$ cannot be measured easily in real
observations. Only the quantity $\sigma^2_{0ran}$ can be estimated. Therefore
observationally defined signal-to-noise ratio is often with respect to $\sigma_{0ran}$ 
rather than to the true noise variance $\sigma_{0}$. For a given 
$\nu_{ran}$, the true signal-to-noise ratio $\nu$ 
decreases with the increase of $\sigma_{0corr}$. Because $N_{peak}(\nu)$ drops steeply at 
large $\nu$, the cumulative number of false peaks given a threshold on $\nu_{ran}$ increases
sensitively as $\sigma_{0corr}$ increases. This can result a large reduction of the efficiency 
of weak lensing cluster detections. If a $75\%$ efficiency is expected 
in the case $\sigma_{0corr}=0$ for a survey at a detection threshold $\nu_{ran}=3.5$, 
this number goes down to $65\%$ with $\sigma^2_{0corr}\approx 0.1\sigma^2_{0ran}$.
The increase of the number of false peaks with intrinsic alignments can also affect
the statistical likelihood in judging whether a dark peak truly corresponds to a dark clump
or is a false one from intrinsic ellipticities of source galaxies. 

On the other hand, the number of false peaks can be a sensitive probe to the
intrinsic alignments of background galaxies. A value $\sigma^2_{0corr}\approx 0.1\sigma^2_{0ran}$
results $N_{peak}(\nu_{ran}\ge 3.5, \sigma_{0corr})=1.7N_{peak}(\nu_{ran}\ge 3.5, \sigma_{0corr}=0)$.
Thus it is easier to derive information of $\sigma^2_{0corr}$ from $N_{peak}(\nu_{ran})$ 
than to directly measure $\sigma^2_{0corr}$ from two-point correlations. The studies 
of CFHTLS Deep cluster survey find that $5$ out of $14$ peaks with $\nu_{ran}>3.5$
in the area of $3.61\hbox{ deg}^2$ are possibly false ones. We then obtain a constraint 
$\sigma^2_{0corr}<2\times 10^{-5}$ ($1\sigma$), which corresponds to the constraint on 
$A$ in eq. (4) with $A< 2.9$, fully consistent with the limit from SDSS observations.
Future large surveys can generate samples containing many lensing detected 
candidates. If one can find $50$ false peaks, the quantity $\sigma^2_{0corr}/\sigma^2_{0ran}$
can be constrained to $\delta (\sigma^2_{0corr}/\sigma^2_{0ran})\sim 1.8\%$. 
With $\sigma^2_{0ran}\sim 4\times 10^{-4}$,
we then have $\delta (\sigma^2_{0corr})\sim 7\times 10^{-6}$ and $\delta (C_{11}+C_{22})
\sim 1.4\times 10^{-5}$.
%

The intrinsic alignments of galaxies carry important information of galaxy formation, 
especially the environmental effects. Previous observational studies show that the 
intrinsic alignments of galaxies are at the lower end of the theoretical predictions 
for dark matter halos, indicating the possible existence of misalignment between galaxies 
and their host halos (e.g., Heymans et al. 2006). 
The method proposed in this paper allows us to constrain the 
intrinsic alignments of galaxies to a very high precision, and therefore is very 
promising in detailed studies on the formation of galaxies. 

In our analysis, we assume a Gaussian statistics for the noise in smoothed $\kappa$ fields
for both cases with or without intrinsic alignments. 
Although we do not expect a highly non-Gaussian smoothed noise field
because of the central limit theorem,
its detailed statistical properties deserve thorough investigations. 
The existence of intrinsic ellipticities as well
as their alignments not only produces false peaks in $\kappa$ maps, but also affects the 
height of the true peaks that are associated with clusters of galaxies. From 
previous discussions, we have $K_N(\vec \theta)=K(\vec \theta)+N(\vec \theta)$ for smoothed
$\kappa$ fields. Assuming that $K$ and $N$ are independent of each other, we can write
the distribution of $K_N$ in the form $p(K_N)dK_N=[\int p_{K}(K)p_{N}(K_N-K)dK]dK_N$.
It is seen that $p(K_N)$ depends on the statistics of $K$ and $N$. Thus detailed
analyses on the statistical properties of $K$ and $N$ are crucial in order to understand 
how the true peaks are influenced by the noise. Further complications
arise due to the shear-ellipticity correlations (e.g., Hirata \& Seljak 2004; Mandelbaum et al. 2006). 
Because the ellipticities of galaxies are associated with the properties of their host halos, 
correlations between the ellipticities of foreground galaxies and the shears generated by 
their host halos on background galaxies are expected. Then $K$ and $N$ are not 
independent quantities anymore. The effects of intrinsic alignments and
shear-ellipticity correlations on lensing analyses depend differently on the redshift distribution of 
background galaxies. The narrower the distribution is, the stronger
the effects of the intrinsic alignments are. For shear-ellipticity correlations, 
the effects are stronger for broader distributions. For tomographic lensing studies,
the intrinsic alignments are important for galaxies within the same redshift bins,
while the shear-ellipticity correlations are significant in considering the cross correlations
between different bins. Extensive investigations on these problems and their
effects on weak lensing cluster surveys will be pursued in our future research. 

Weak lensing cluster studies together with other lensing analyses are sensitive probes of
the dark matter distribution as well as the nature of dark energy (e.g., Fang \& Haiman 2006). 
With fast observational advances and thorough theoretical understandings of different systematics,
cosmological applications of weak lensing effects will greatly improve our knowledge 
about the universe.

\acknowledgments 
We sincerely thank the referee for the encouraging and constructive comments and 
suggestions. This research was supported in part by the
National Science Foundation of China under grants 10243006 and
10373001, by the Ministry of Science and Technology of China
under grant TG1999075401, by the Key Grant
Project of Chinese Ministry of Education (No. 305001),
and by the National Science Foundation of China under grant 10533010.


\begin{figure}
\plotone{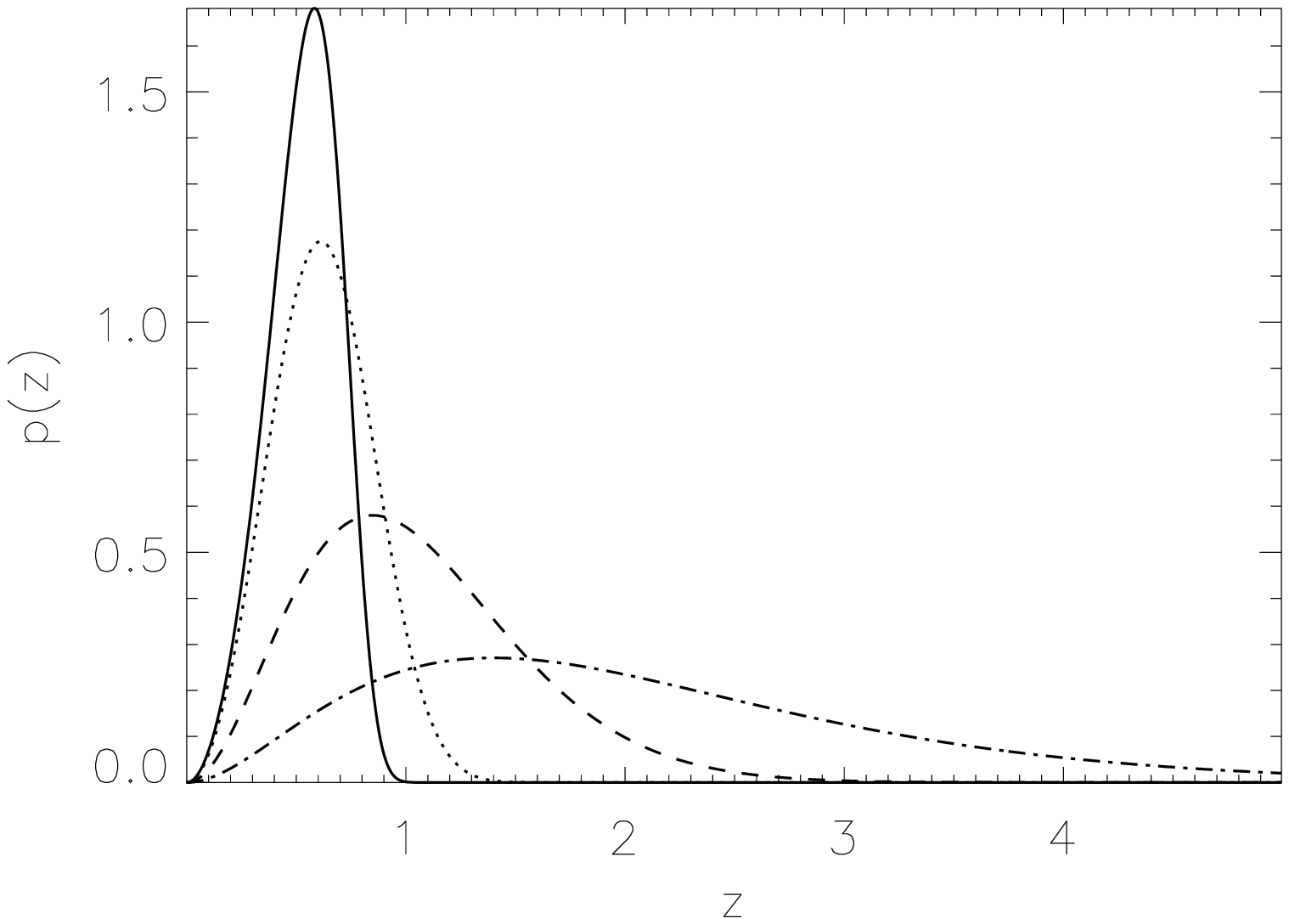}
\caption{The redshift distribution of background galaxies with the functional form
given in eq. (24). Here we take $\alpha=2$ and $z_s=0.7$. The solid, dotted, dashed, and
dash-dotted lines correspond to $\beta=6$, $3$, $1.5$ and $1$, respectively.
\label{yg6}}
\end{figure}

\begin{figure}
\plotone{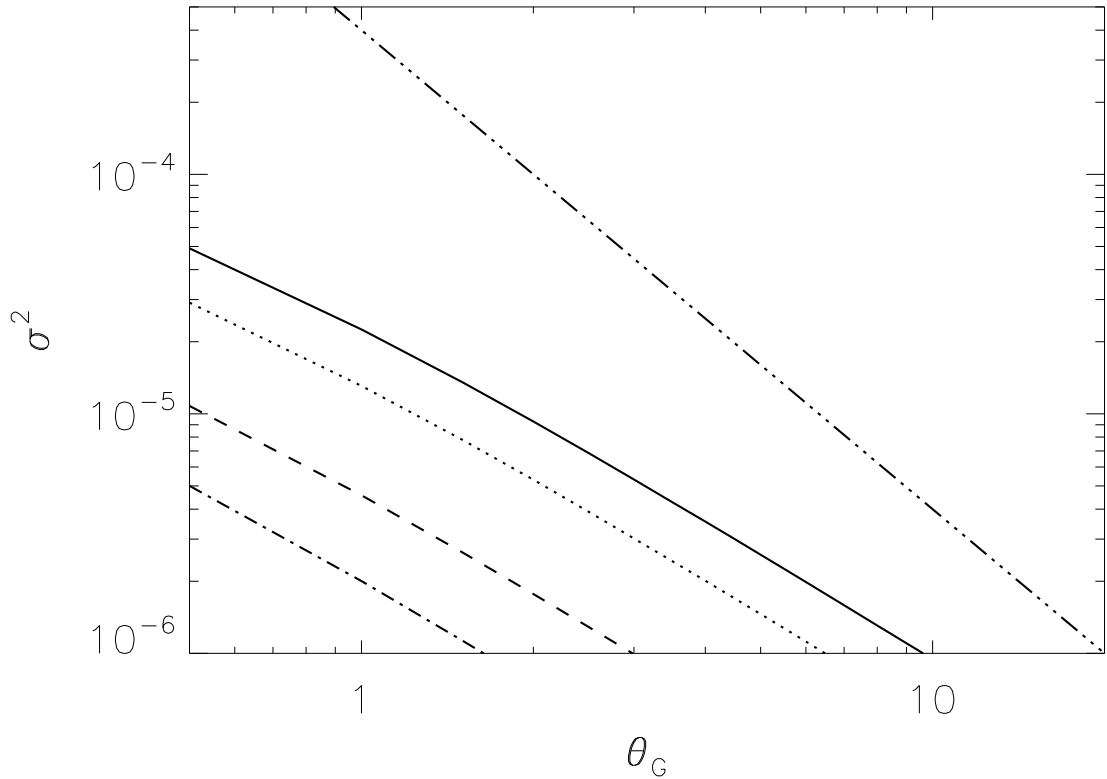}
\caption{The variance contributed by intrinsic alignments. We take $A=0.57$ in eq. (4),
and $\alpha=2$ and $z_s=0.7$ in eq. (24). The solid, dotted, dashed, and
dash-dotted lines correspond to the results with $\beta=6, 3, 1.5$ and $1$, respectively.
The dash-dot-dot-dotted line is $\sigma^2_{0ran}$ with $\sigma_{\epsilon}=0.4$ and
$n_g=30\hbox{ arcmin}^{-2}$.
\label{yg4}}
\end{figure}

\begin{figure}
\plotone{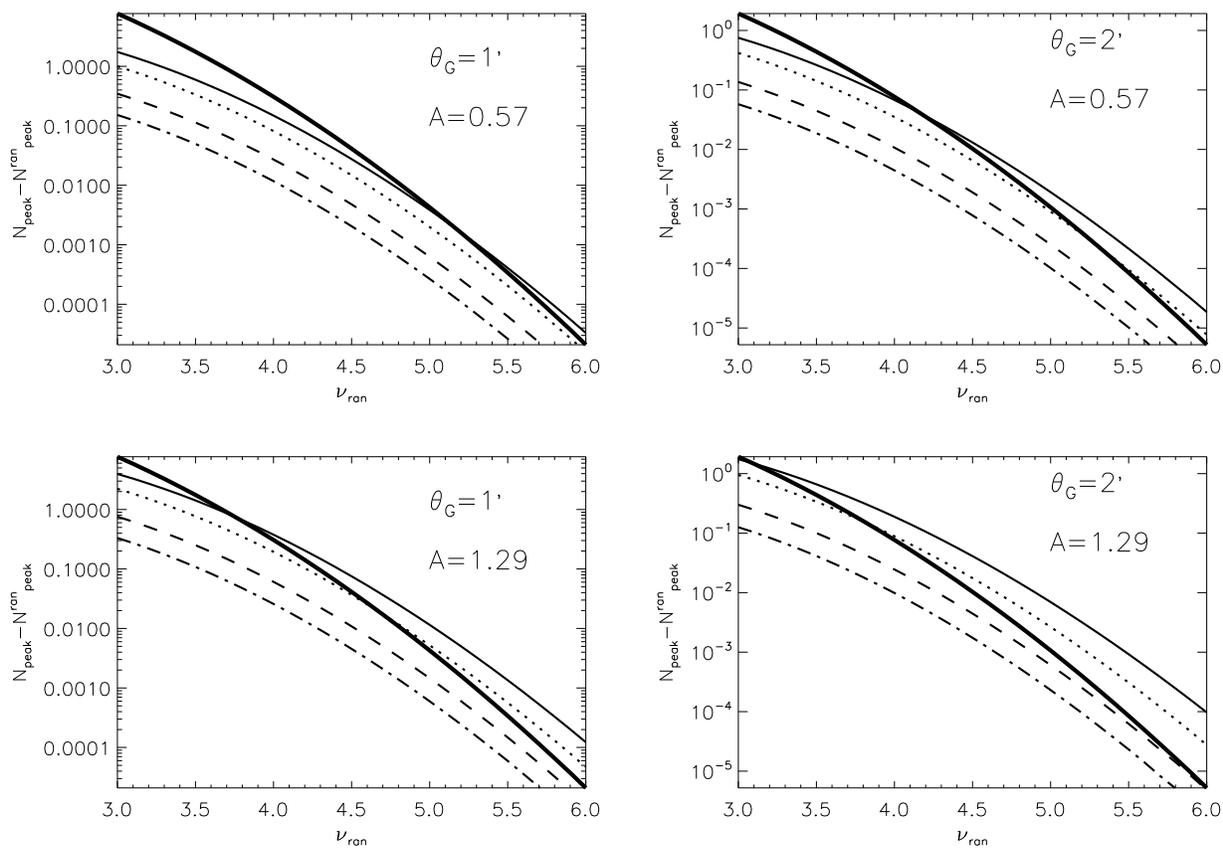}
\caption{The cumulative number of false peaks resulting from intrinsic alignments in $1\hbox{ deg}^2$.
The thick solid line is the cumulative number of false peaks without intrinsic alignment.
The thin solid, dotted, dashed, and dash-dotted lines are for $\beta=6$, $3$, $1.5$, and $1$, 
respectively.
\label{yg7}}
\end{figure}

\begin{figure}
\plotone{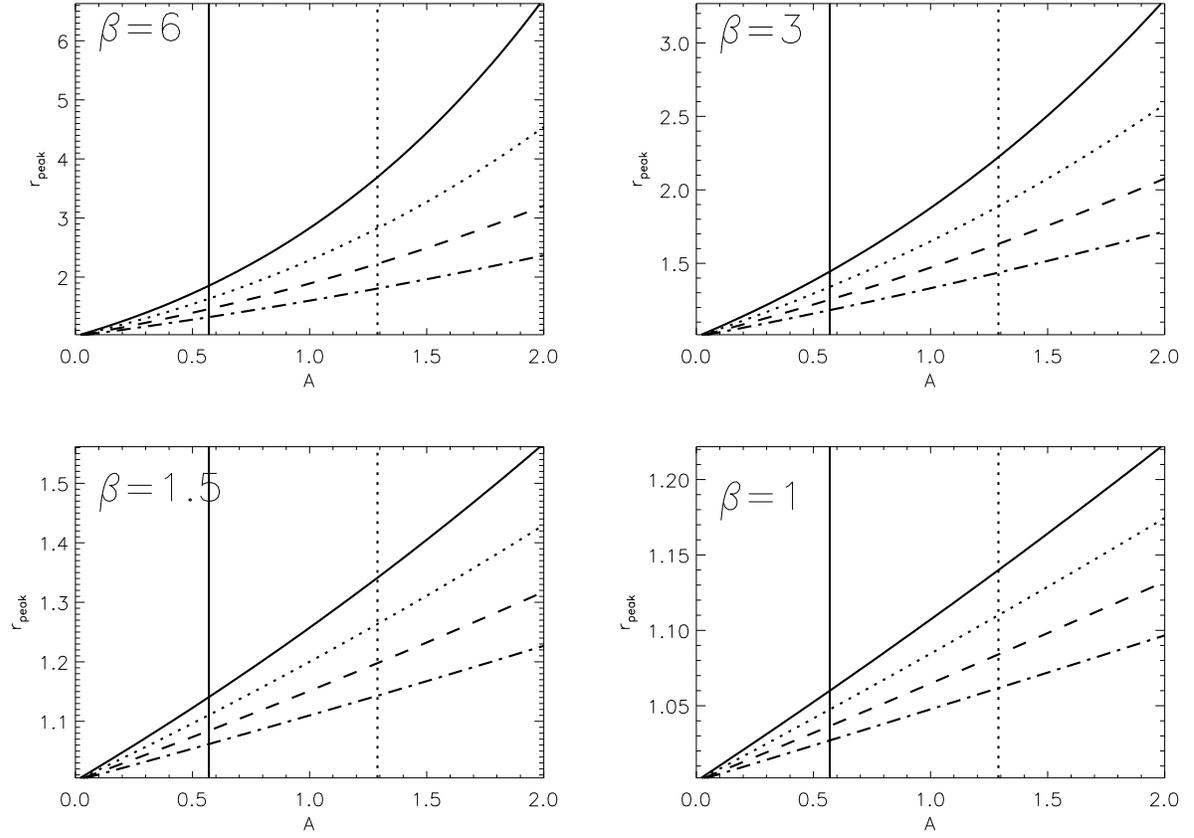}
\caption{The dependence of the ratio $r_{peak}=N_{peak}/N^{ran}_{peak}$ on $A$ for
$\theta_G=1\hbox{ arcmin}$. Different panels show the results with different $\beta$ values.
The solid, dotted, dashed, and dash-dotted lines are for $\nu_{ran}=5$, $4.5$, $4$, and
$3.5$, respectively. The solid and dotted vertical lines are respectively at the position of $A=0.57$ 
and $A=1.29$. 
\label{yg8}}
\end{figure}



\begin{figure}
\plotone{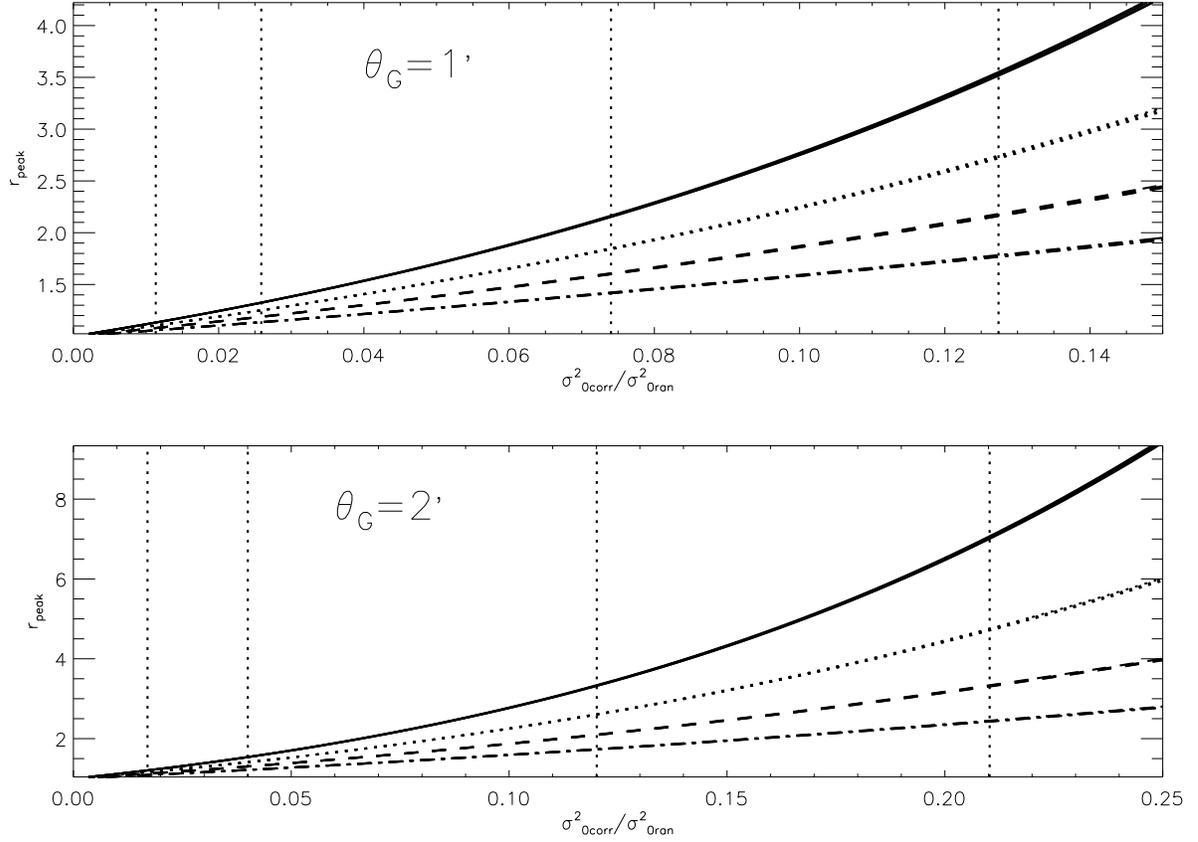}
\caption{The ratio $r_{peak}$ with respect to $\sigma^2_{0corr}/\sigma^2_{0ran}$. The upper
and lower panels correspond to $\theta_G=1\hbox{ arcmin}$ and $\theta_G=2\hbox{ arcmin}$, 
respectively. The four sets of solid, dotted, dashed and dash-dotted lines
are respectively for $\nu_{ran}=5$, $4.5$, $4$, and $3.5$. Each set of lines contains
results with $\beta=6$, $3$, $1.5$ and $1$. The vertical dotted lines from right to
left show the value of $\sigma^2_{0corr}/\sigma^2_{0ran}$ with $A=1.29$ 
for $\beta=6$, $3$, $1.5$ and $1$, respectively.
\label{yg2}}
\end{figure}

\begin{figure}
\plotone{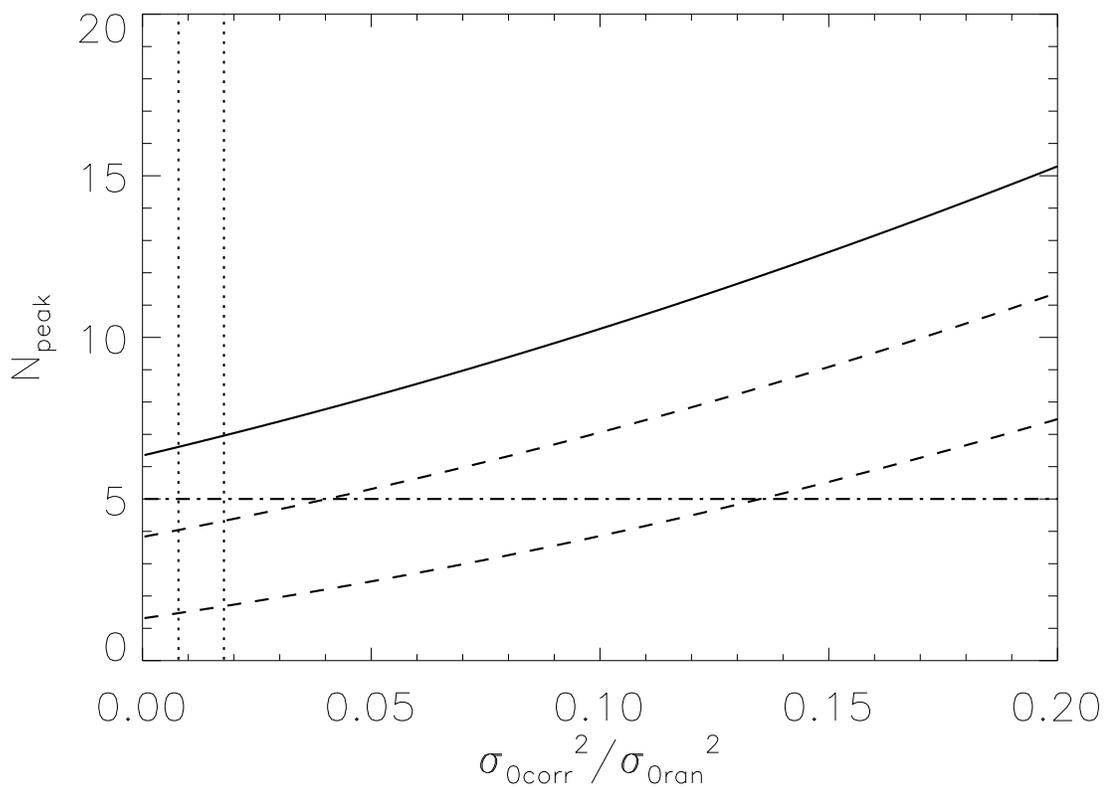}
\caption{The number of false peaks with $\nu_{ran}\ge 3.5$
in $3.61\hbox{ deg}^2$. 
The solid line is for the average number of false peaks.
The upper and lower dashed lines represent $-1\sigma$
and $-2\sigma$ Poisson deviations from the mean, respectively. The horizontal dash-dotted line is
at $N_{peak}=5$. The two vertical dotted lines correspond to the values of
$\sigma^2_{0corr}/\sigma^2_{0ran}$ for the CFHTLS Deep with $A=0.57$ and $A=1.29$, from
left to right, respectively.
\label{yg2}}
\end{figure}

\begin{table}
\caption{Ratio of $\sigma^2_{0corr}/\sigma^2_{0ran}$ with 
$\alpha=2$, $z_s=0.7$, $n_g=30\hbox{ arcmin}^{-2}$ and $\sigma_{\epsilon}=0.4$}\label{tab1}
\begin{center}
\begin{tabular}{llllllllllllll}\hline
  $\beta$ & $z_{med}$ & $\sigma^2_{0corr}/\sigma^2_{0ran}$ &
$\sigma^2_{0corr}/\sigma^2_{0ran}$ & $\sigma^2_{0corr}/\sigma^2_{0ran}$ & $\sigma^2_{0corr}/\sigma^2_{0ran} $ & \\
&  & $(\theta_G=1^{'})$ & $(\theta_G=2^{'})$ & $(\theta_G=1^{'})$ & $(\theta_G=2^{'})$ & \\
&  & $(A=0.57)$ & $(A=0.57)$ & $(A=1.29)$ & $(A=1.29) $ & \\
 \hline\hline
      &  &  &  &  &  \\
  $6$    & $0.55$ & $6\%$ & $10\%$ & $13.5\%$ & $22.6\%$ \\
      &  &  &  &  &  \\
  $3$    & $0.62$ & $3\%$ & $5\%$ & $6.8\%$ & $11.3\%$ \\
      &  &  &  &  & \\
  $1.5$  & $0.99$ & $1\%$ & $2\%$ & $2.3\%$ & $4.5\%$ \\
    &  &  &  &  &  \\
  $1$    & $1.87$ & $0.5\%$ & $0.8\%$ & $1.1\%$ & $1.8\%$ \\
    &  &  &  &  &  \\
 \hline
 \end{tabular}
\end{center}
\end{table}

\begin{table}
\caption{Ratio of $\sigma^2_{0corr}/\sigma^2_{0ran}$ for different surveys. 
For the redshift distribution, we take $\alpha=2$ and $\beta=1.5$.
}\label{tab1}
\begin{center}
\begin{tabular}{llllllllllllll}\hline
 & $<z>$ & $z_s$ & $z_{med}$ & $n_g $ & $\sigma^2_{0corr}/\sigma^2_{0ran}$ &
$\sigma^2_{0corr}/\sigma^2_{0ran}$ & \\
& &  &  & $(\hbox{ arcmin}^{-2})$  & $(\theta_G=1^{'})$ & $(\theta_G=2^{'})$ & \\
& &  &  &  & $(A=0.57)$ & $(A=0.57)$ & \\
 \hline\hline
   &   &  &  &  &  &  \\
  COSMOS & $1.2$   & $0.8$ & $1.13$ & $70$ & $2.3\%$ & $3.5\%$ \\
   &   &  &  &  &  &  \\
  SNAP &   $1.2$   & $0.8$ & $1.13$ & $100$ & $3.3\%$ & $5\%$ \\
   &   &  &  &  &  & \\
  SNAP(Deep) & $1.4$  & $0.93$ & $1.3$ & $300$ & $8.4\%$ & $12.8\%$ \\
    &  &  &  &  &  & \\
    &  &  &  &  &  & \\
 \hline
 \end{tabular}
\end{center}
\end{table}

\begin{table}
\caption{Probability for the detected clumps being false peaks from chance alignments 
of background galaxies. Here $n_{clump}$ denotes the number of clumps found in the field,
$p(ran)$ represents the probability without intrinsic alignments, and
$p(corr)$ is the probability taking into account intrinsic alignments with $A=1.29$.}\label{tab1}
\begin{center}
\begin{tabular}{llllllllllllll}\hline
Obs. & Area &  $n_g$ & $\theta_G$ & $\nu_{ran}$ & $n_{clump}$
& $p(ran)$ & $p(corr)$  \\
& $\hbox{deg}^2$ & $(\hbox{ arcmin}^{-2})$ &  $(\hbox{arcmin})$ &  &  & & \\
& &  &  &  &  &  &\\
 \hline\hline
   &   &  &  &  &  & & \\
  Erben et al. & $0.05$ & $20$ & $0.5$ & $4.5$ & $1$ & $0.9\%$ & $1\%$ \\
   (2000) &   &  &  &  &  & & \\
  Linden et al. & $0.05$ & $65$ & $0.5$ & $3$ & $1$ & $30\%$  & $27\%$ \\
   (2006) &   &  &  &  &  & & &\\
  Massey et al. & $2$  & $70$ & $1$ & $4$ & $2$ & $10\%$ & $15\%$ \\
   (2007)&  &  &  &  &  & & \\
  Massey et al. & $2$  & $70$ & $1$ & $4.5$ & $2$ & $0.3\%$ & $0.6\%$ \\
   (2007)&  &  &  &  &  & & \\
  Massey et al. & $2$  & $70$ & $1$ & $5$ & $2$ & $0.003\%$ & $0.01\%$ \\
   (2007)&  &  &  &  &  & & \\
    &  &  &  &  &  & &\\
 \hline
 \end{tabular}
\end{center}
\end{table}





\begin{thebibliography}{}
   \bibitem[Bartelmann \& Schneider (2001)]{bs01}Bartelmann, M., \& Schneider, P. 2001, Physics Reports, 340, 291
   \bibitem[Bode(2006)]{}Bode, P., Ostriker, J. P., Weller, J., \& Shaw, L. astro-ph/0612663
   \bibitem[Bond(1987)]{}Bond, J. R, \& Efstathiou, G. 1987, \mnras, 226, 655
   \bibitem[Borgani(2006)]{}Borgani, S. 2006, astro-ph/0605575 
   \bibitem[Brown(2002)]{}Brown, M. L., Taylor, A. N., Hambly, N. C., \& Dye, S. 2002, \mnras, 333, 501
   \bibitem[Croft(2000)]{}Croft, R. A. C., \& Metzler, C. A. 2000, \apj, 545, 561
   \bibitem[Crittenden(2001)]{}Crittenden, R. G., Natarajan, P., Pen, U. L., \& Theuns, T. 2001, 
\apj, 559, 552
   \bibitem[Erben(2000)]{}Erben, Th., van Waerbeke, L., Mellier, Y., Schneider, P., Cuillandei, J. C., Castander, F. J. \& Dantel-Fort, M. 2000, A\&A, 355, 23
   \bibitem[Fang(2006)]{}Fang, W., \& Haiman, Z. 2006, astro-ph/0612187
   \bibitem[Gavazzi(2007)]{}Gavazzi, R., \& Soucail, G. 2007, A\& A, 462, 259
   \bibitem[Hamana(2004)]{}Hamana, T., Takada, M., \& Yoshida, N. 2004, \mnras, 350, 893 
   \bibitem[Heavens(2000)]{}Heavens, A., Refregier, A., \& Heymens, C. 2000, \mnras, 319, 649
   \bibitem[Hennawi(2005)]{}Hennawi, J. F. \& Spergel, D. N. 2005, \apj, 624, 59
   \bibitem[Heymans(2004)]{}Heymans, C., Brown, M., Heavens, A., Meisenheimer, K. Taylor, A., \& Wolf, C. 2004, \mnras, 347, 895
   \bibitem[Heymans(2006)]{}Heymans, C., White, M., Heavens, A., Vale, C., \& van Waerbeke, L.  2006, \mnras, 371, 750
   \bibitem[Hirata(2004)]{}Hirata, C. M., \& Seljak, U. 2004, Phys. Rev. D., 70, 063526
   \bibitem[Hoekstra(2006)]{}Hoekstra, H., Mellier, Y., van Waerbeke, L., Semboloni, E., Fu, L., Hudson, M. J., Parker, L. C., Tereno, I., \& Benabed, K. 2006, \apj, 647, 116
   \bibitem[Jing(2002)]{}Jing, Y. P. 2002, \mnras, 335, L89
   \bibitem[Kaiser(1993)]{}Kaiser, N., \& Squires, G. 1993, \apj, 404 441
   \bibitem[King(2003)]{}King, L. J., \& Schneider, P. 2003, A\&A, 398, 23
   \bibitem[King(2006)]{}King, L. J. 2006, A\&A, 441, 47 
   \bibitem[Knox(2006)]{}Knox, L. , Song, Y., \& Tyson, J. A. 2006, Phys. Rev. D., 74, 023512
   \bibitem[Linden(2006)]{}von der Linden, A., Erben, T., Schneider, P., \& Castander, F. J. 2006, A\&A, 454, 37 
   \bibitem[Mandelbaum(2006)]{}Mandelbaum, R. , Hirata, C. M., Ishak, M., Seljak, U., \& Brinkmann, J. 2006, \mnras, 367, 611
   \bibitem[Massey(2007)]{}Massey, R. et al. 2007, Nature, 445, 286 
   \bibitem[Munshi(2006)]{}Munshi, D. , Valageas, P., van Waerbeke, L., \& Heavens, A. 2006, astro-ph/0612667 
   \bibitem[NFW(1996)]{}Navarro, J., Frenk, C. \& White, S. D. M. 1996, \apj, 462, 563
   \bibitem[Porciani(2002)]{}Porciani, C. , Dekel, A., \& Hoffman, Y. 2002, \mnras, 332, 32
   \bibitem[Semboloni(2007)]{}Semboloni, E., van Waerbeke, L., Heymans, C., Hamana, T., Colombi, S., White, M., \& Mellier, Y. 2007, \mnras, 375L, 6
   \bibitem[Schirmer(2007)]{}Schirmer, M., Erben, T., Hetterscheidt, M., \&  Schneider, P. 2007, A\& A, 462, 875
   \bibitem[Schneider(1996)]{}Schneider, P. 1996, \mnras, 283, 837
   \bibitem[Starck(2006)]{}Starck, J. L., Pires, S. \& Refregier, A. 2006, A\& A, 451, 1139
   \bibitem[Tang(2005)]{}Tang, J. Y., \& Fan, Z. H. 2005, \apj, 635, 60 
   \bibitem[Waerbeke(2000)]{}van Waerbeke, L. 2000, \mnras, 313, 524 
   \bibitem[White(2002)]{}White, M., van Waerbeke, L., \& Mackey, J. 2002, \apj, 575, 640
\end{thebibliography}
\end{document}